\documentclass[aps,pra,twocolumn]{revtex4-1}
\usepackage{graphicx}
\usepackage{epstopdf}
\usepackage{color}
\usepackage{amsmath}
\usepackage{float}
\usepackage{latexsym}
\usepackage{natbib}
\usepackage{ amssymb }
\DeclareMathOperator{\Tr}{Tr}

\begin{document}

\title{A Method of Developing Analytical Multipartite Delocalization Measures for Mixed W-like States}

\author{Cathal Smyth}
 \affiliation{Department of Physics, University of Toronto, Toronto, Canada}
\author{Gregory D. Scholes}
 \email{gscholes@princeton.edu}
 \affiliation{Department of Chemistry, University of Toronto, Toronto, Canada}
 \affiliation{Department of Chemistry, Princeton University, Washington Rd, Princeton NJ 08544, USA}

\begin{abstract}

\noindent We present a method of developing analytical measures of $k$-partite delocalization in arbitrary $n$-body W-like states, otherwise known as mixed states in the single excitation subspace. These measures calculate the distance of a state to its closest reference state with $k-1$ entanglement. We find that the reference state is determined by the purity of the state undergoing measurement. Measures with up to 6-body delocalization for a 6-body system are derived in full, while an algorithm for general $k$-partite measures is given.

\end{abstract}

\maketitle
\section{Introduction}

\noindent  The role of entanglement has extended far beyond fundamental quantum mechanics to fields as diverse as quantum computing \citep{Nielsen2000}, astrophysics \citep{Gottesman2004} and now energy transfer in photosynthetic systems, where the concepts of entanglement are employed in measuring delocalization of electronic excitation among light-absorbing molecules \citep{Smyth2012,Fassioli2010a,Sarovar2010,Caruso2010,Scholak2011,Levi2014,Scholes2014}. While earlier work has centered on simple measures of wavefunction delocalization \citep{Thouless1974} that neglect homogeneous line broadening (dephasing), such entanglement based statistical measures have proven to be much more powerful assets to the field of energy transfer \citep{Smyth2012}. We consider an interesting question from the quantum mechanical viewpoint: Precisely how is an excitation shared among chromophores and how can that be characterized? In this paper we develop a method for determining $k$-partite delocalization ($k\leq n$) in $n$-mode open systems with only one excitation. With the restriction of a single excitation comes the added benefit that the degree of delocalization of the excitation can also be viewed as mode entanglement; indeed in this paper we will use the two concepts interchangeably. This method allows us to measure $k$-partite delocalization by taking advantage of the tiered structure of separability in $k$-partite entanglement.\\

% The role of entanglement has extended far beyond fundamental quantum mechanics to fields as diverse as quantum computing \citep{Nielsen2000}, astrophysics \citep{Gottesman2004} and even energy transfer in photosynthetic systems \citep{Smyth2012,Fassioli2010a,Sarovar2010,Caruso2010}.
\noindent Qualifying and quantifying the presence of entanglement has proven an arduous task, often increasing exponentially with the number of parties. In the multipartite setting, notions such as ``maximal entanglement", and separability are no longer black and white; inequivalent classes of entanglement such as the GHZ \citep{Greenberger1990} and W-states \citep{Dur2000} arise, while states form a tiered structure of separability. The difficulty of detecting entanglement is further compounded when considering open systems, i.e. systems that undergo decoherence \citep{Carvalho2004,Huber2010}. For a review, see \citet{Horodecki2009}, \citet{Mintert2005}, \citet{Plenio2007} or \citet{Guhne2008}.\\

\noindent While focusing solely on the single excitation subspace greatly reduces the amount of information to be processed, it should be noted that states subject to these measures must be formed and preserved within this subspace. Under any local operations that preserve the state within this subspace, the measures described here can be considered to accurately detect multipartite entanglement. The presence of populations in other subspaces will reduce or even remove such entanglement \citep{Tiersch2012}, and we caution that truncating such subspaces can lead to false detection of entanglement. However, as measures of delocalization our functions will still be effective, because the tiered structure of coherence remains intact. As such, the functions described here are emphatically not entanglement measures as they are not invariant under local transformations. Nevertheless, for a system purely within the single excitation subspace, we have proposed a systematic method of detecting multipartite entanglement in mixed W-like states.  In quantum information, a state with a single excitation shared across $n\geq 3$ modes is known as a W-like state. W-states are vital in quantum information theory as they are robust against decoherence \citep{Carvalho2004}, and may provide a valuable resource for scalable quantum information processing. Therefore, quantifying the entanglement of such states is essential. W-states can be produced in experiments involving atomic ensembles \citep{Choi2010}, as well as single photon entanglement \citep{Papp2009}. We believe that the functions described in this paper could, with modification, lead to such quantification.
\\

\noindent A famous example of entanglement measures is the relative entropy \citep{Vedral1997}. This measure compares the entropy of a state to its closest separable state. In contrast, our delocalization measures make use of the tiered structure of separability in multipartite entanglement and, in order to quantify $k$-partite delocalization, compare a state to its nearest $(k-1)$-partite entangled state. Some of the criteria employed are analogous to those within other approaches in the literature; \citet{Papp2009} considered entanglement detection as a function of the degree of photon contamination (in our case as a function of purity), while \citet{Blasone2008} looked at creating distance measures for $k$-partite entangled pure states by measuring the distance from the closest $(k-1)$-partite entangled state.
\\
\section{Delocalization measures}
 \noindent In the single excitation subspace, a convenient equivalence between coherence and entanglement arises \citep{Sarovar2010}. A measure of bipartite entanglement, the tangle \citep{Coffman2000}, is related to coherence between modes \emph{a} and \emph{b} by ${\tau }_{ab}=4{\left|{\rho }_{ab}\right|}^2 $.
Adding all the possible tangles gives the total tangle, or total bipartite entanglement in the system \citep{Fassioli2010a}
\begin{equation}
E_2\left(\rho \right)=\sum^N_{a=1,b\neq a}{{\tau }_{ab}}.
\label{bound}
  \end{equation}

 \noindent This measure can be rewritten as a function of the purity of the state, and the second order statistical moment \citep{Smyth2012}

  \begin{equation} E_2\left(\rho \right)=\Tr\left({\rho }^2\right)-M_2\left({\rho }\right).
  \label{T2}
  \end{equation}

\noindent This statistical measure $M_2\left({\rho }\right)$ is also known as the Inverse Participation Ratio, a measure of bipartite delocalization in pure states \citep{Thouless1974}. Unlike the purity measure, it is basis dependent, and should be applied to the basis under investigation. In this paper, we focus on the single excitation subspace of the computational basis. In general a statistical moment of order $k$ is written as   \begin{equation}
M_{k}\left(\rho\right)=\sum^N_{j=1}\left(\rho_{jj}\right)^{k}.
  \end{equation}
\\

\noindent Measures of multipartite delocalization in pure W-like states have already been developed \citep{Scholak2011}. These measures make use of the statistical moments of the state populations to detect and quantify $k$-partite entanglement in $n$-body systems. The equations for bipartite up to quinquepartite entanglement are given below.
\begin{widetext}
\begin{subequations}
\label{allequations}
\begin{eqnarray}
\tau_{2}\left(\rho\right)&=&1-M_{2}\left({\rho}\right),\label{equationa} \\
\tau_{3}\left(\rho\right)&=&1-3{M_{2}\left(\rho\right)}+2{M_{3}\left(\rho\right)},\label{equationb} \\
\tau_{4}\left(\rho\right)&=&1-6M_{2}\left({\rho}\right)+8M_{3}\left({\rho}\right)+{3M_{2}\left({\rho}\right)}^2-6M_{4}\left({\rho}\right),\label{equationc}\\
\tau_{5}\left(\rho\right)&=&1-10M_{2}\left({\rho}\right)+20M_{3}\left({\rho}\right)+15M_{2}\left({\rho}\right)^2-30M_{4}\left({\rho}\right)-20M_{2}\left({\rho}\right)M_{3}\left({\rho}\right)+24M_{5}\left({\rho}\right)\label{equationd}\\
\nonumber
\end{eqnarray}
\end{subequations}
\end{widetext}
\subsection{Bipartite delocalization measures}
\noindent  It is evident that equations \ref{T2} and \ref{equationa} are equivalent in the case of a pure state. Our objective is to derive mixed state versions of equations \ref{equationb}-\ref{equationd} that reduce to the pure state equations when $\Tr\left({\rho }^2\right)=1$. We can begin by reinterpreting equation \ref{T2} as a measure of the distance of state $\rho$ from its nearest separable state $\sigma$ such that
  \begin{figure}[htbp]

  % <PostScript inclusion special>
\begin{center}
\includegraphics[scale=0.45]{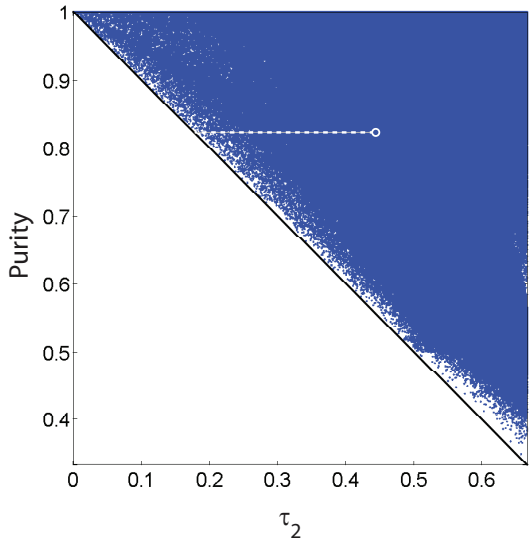}

\end{center}
\caption{
Distribution of $\tau_{2}$  versus purity for a set of 1 million randomly generated 3 body states (blue, upper right corner) and 1 million randomly generated reference (separable) states (black diagonal line). The closest reference states are those with the same level of purity. Overlayed is a small circle indicating an example state, with a dashed line leading to its nearest separable state.}
  \label{E2plot}
\end{figure}
 \begin{equation}
E_2\left(\rho \right)=\tau_{2}\left(\rho\right)-\tau_{2}\left(\sigma\right).
\label{E2distance}
 \end{equation}

\noindent In this case, $\sigma$ is just a diagonal density matrix composed of a distribution of values that is equivalent to the distribution of eigenvalues of $\rho$. Note that $\sigma$ is in the same basis of investigation as $\rho$. Given that there are no off-diagonal elements in $\sigma$, it has no entanglement, and also happens to have the same purity as $\rho$. When a matrix like $\sigma$ is diagonal, its purity $\Tr\left({\sigma}^2\right)$ and $M_2\left({\sigma}\right)$ are equivalent, and thus so are equations \ref{T2} and \ref{E2distance}. In other words, $\sigma$ minimizes the distance between the entangled and separable regimes. To illustrate this point, the purity and measure $\tau_2$ of 1 million randomly generated 3 body states and 1 million randomly generated reference (separable) states are plotted in figure \ref{E2plot}. Here we can see that for any given purity there is only possible reference state, which is also the closest separable state. We can now expand upon the idea of measuring distance within the context of tripartite entanglement. \\

\subsection{Tripartite delocalization measures}
\noindent Just like in the bipartite measure, we will apply a measure (in this case equation \ref{equationb}) to our state $\rho$, as well as some reference state $\sigma$, of equal purity:
 \begin{equation}E_{3}\left(\rho\right)\equiv \tau_{3}\left(\rho\right)-\tau_{3}\left(\sigma\right).\end{equation}%\Tr\left(\rho^2\right)
%where the prefactor $ \Tr\left(\rho^2\right)$ is added to ensure convexity (see supplementary material).
  \begin{figure}[htbp]

  % <PostScript inclusion special>
\begin{center}
\includegraphics[scale=0.45]{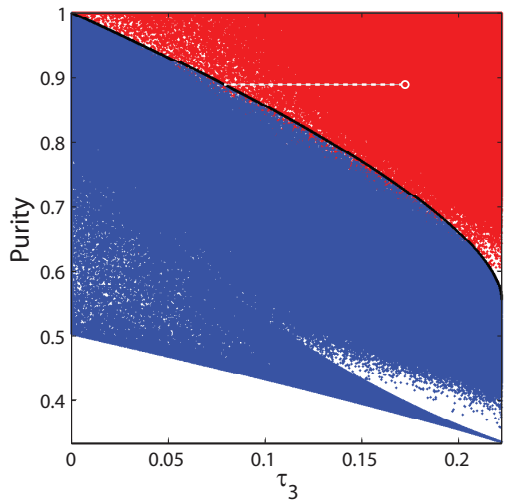}

\end{center}
\caption{
Distribution of $\tau_{3}$  versus purity for a set of 1 million randomly generated 3 body states (red, upper right corner) and 1 million randomly generated reference (biseparable) states (blue, centre). The closest reference states lie along the border between these two regions (denoted by a black curve), running from $\textrm{Purity}=1$ down to $\textrm{Purity}=5/9$. Overlayed is a small circle indicating an example state, with a dashed line leading to its nearest biseparable state.}
  \label{E3plot}
\end{figure}

 \noindent The requirement that our state $\sigma$ should have the same purity as $\rho$ makes sense. Not only is it a natural extension of our bipartite measure but also purity is a measure of entropy. As states become more mixed they become less distinguishable; for a given level of entropy a reference point is needed to distinguish our states. In figure \ref{E3plot} we plot the purity and $\tau_3$ for $1$ million randomly generated $3$-body states and $1$ million random bipartite states. As can be seen in figure \ref{E3plot}, the reference states generated form a distinct convex-shaped border with the $3$-body states; meaning that, for a given level of purity, only one reference state $\sigma$ can play the role of closest state to $\rho$. Unlike the bipartite measure however, the state $\sigma$ will not be a separable state, but rather is defined as the closest state with bipartite entanglement. This ensures we can distinguish our state from others that are bipartite entangled but not tripartite entangled. The problem now falls to finding the closest bipartite state.
 \\

 \noindent The higher the level of bipartite entanglement, the higher the likelihood that there are higher orders of entanglement. We can see this by imagining infinite-body pure states with varying levels of entanglement. A state with at most 2-body entanglement can have a maximum value of $E_2\left(\rho\right)=1/2$, a state with at most 3-body entanglement $E_2\left(\rho\right)=2/3$, and so on as the value of $E_2\left(\rho\right)$ approaches $1$ as $k$ approaches infinity. As our states become mixed, their maximal value for  $E_2\left(\rho\right)$  decreases as a function of the purity. In order to minimize the distance between your state of interest $\rho$ and the reference state $\sigma$, the bipartite entanglement of $\sigma$ must be large; meaning that, according to equation \ref{T2}, one needs a small value of $M_2\left(\sigma\right)$ for a given level of purity. In general, $\sigma$ is comprised of probability-weighted pure states  $\sigma=p_1\sigma_{1}+p_2\sigma_{2}+\ldots p_n\sigma_{n}$. %, where the probabilities are functions of the state purity and are ordered as $p_1 \geq p_2 \geq \ldots \geq p_n$.
 However, when adding any two density matrices with population overlap, $M_2\left(\sigma\right)$ will not be minimized. A detailed proof follows in the next section.\\

 \noindent Therefore, for a 3-body system, one can envision these states to take on the form (or some permutation thereof):
%\begin{equation}
%\sigma=p\left(\left|\psi^+\right>\left<\psi^+\right|\right)+(1-p)\left|001\right>\left<001\right|
%%\sigma=\frac{p}{{2}}\left(\left|100\right>+\left|010\right>\right)\left(\left<100\right|+\left<010\right|\right)+(1-p)\left|001\right>\left<001\right|
%\end{equation}
%$\left|\psi^+\right>=\frac{1}{\sqrt{2}}(\left|100\right>+\left|010\right>)$ and
  \begin{equation}\sigma=\left( \begin{array}{ccc}
\frac{p}{2} & \frac{p}{2} & 0 \\
\frac{p}{2} & \frac{p}{2} & 0 \\
 0 & 0 & \left(1-p\right)\end{array}
\right)\end{equation}
\\

\noindent where  the probability $p$ is defined as a function of the purity of the state $\rho$
  \begin{equation}
p=\frac{1}{2}\left(1+\sqrt{2\left({\Tr\left({\rho }^2\right)}\right)-1}\right).
\label{prob2ev}
\end{equation}
and is a solution to the quadratic equation $p^2+(1-p)^2=\Tr\left(\rho^2\right)$.
The value $p$ can run from $1/3$ (when $\Tr\left({\rho }^2\right)=\Tr\left({\sigma }^2\right)=5/9$) up to $1$ (when $\Tr\left({\rho }^2\right)=\Tr\left({\sigma }^2\right)=1$). Below a purity of $5/9$ one can no longer distinguish from biseparable states. In the event that a state lies within the biseparable region, its closest biseparable state is set to itself, in order to avoid having false values.
\\
\subsection{Minimizing $M_2\left(\sigma\right)$}
 \noindent Let us now demonstrate how to minimize the value of  $M_2\left(\sigma\right)$. First let us assume that the reference state $\sigma$ is made up of statistically weighted pure states such that $\sigma=p_1\sigma_1+p_2\sigma_2+\ldots+p_n\sigma_n$. Given that $\sigma$ must be maximally $k-1$-partite entangled for its level of purity, it stands to reason that at least one of the constituent states is also maximally $k-1$ entangled. We must also ensure that the statistical moment $M_2(\sigma)$ is minimized, in order to maximize the amount of entanglement. Now let us look at what happens if we add two density matrices $\sigma_1$ and $\sigma_2$ that have population overlap. Assume $\sigma_1$ is delocalized across $k-1$ modes and that $\sigma_2$ is delocalized across $m$ modes where $2 \leq m\leq k-1$. Each density matrix is statistically weighted, by $p_1$ and $p_2$ respectively. Their contribution to $M_2(\sigma)$ will appear as follows:

 \begin{equation}
M_2(p_1\sigma_1+p_2\sigma_2)=\frac{k-2}{(k-1)^2}{p^2_1}+\frac{m-1}{m^2}{p^2_2} + \left(\frac{{p_1}}{k-1}+\frac{{p_2}}{m}\right)^2
\end{equation}

Expanding this out gives us
 \begin{equation}
M_2(p_1\sigma_1+p_2\sigma_2)=\frac{{p^2_1}}{k-1}+\frac{{p^2_2}}{m} + \frac{2{p_1}{p_2}}{m(k-1)}.
\end{equation}

\noindent This result has an extra cross term $\frac{2{p_1}{p_2}}{m(k-1)} $ compared to just the two statistical moments $M_2(p_1\sigma_1)$ and $M_2(p_2\sigma_2)$ added together.\\
Now let us apply an extra restriction: let the size of the system, $n$, be equal to $k-2+m$. What happens if we remove the overlapping population from $\sigma_2$ and place it elsewhere in $\sigma_2$, effectively reducing the size of $\sigma_2$ to $m-1$ modes? Here we find that again, the case with population overlap ($\sigma_2$ with $m$ modes) has a larger value of $M_2(p_1\sigma_1+p_2\sigma_2)$ than the case where $\sigma_2$ has $m-1$ modes, under the condition
 \begin{equation}
p_2 \leq \frac{2(m-1)}{k-1}p_1.
\end{equation}
Given that $p_1 \geq p_2$ and that $m<k$ this will be true for a given range of probabilities. For example, in the measures $E_4$, $E_5$ and $E_6$ defined in the paper, $p_2=1-p_1$. Thus this will always hold true as $\frac{k-1}{2m+k-3} < p_1$. Likewise for $E_3$, $m=k$ and $p_1 \geq p_2=p_3$.
\\
\subsection{General Method }
% In order to do so, we must prove that the constituent pure states of these mixed reference states must not overlap.
\noindent A clear pattern has emerged: In order to measure $k$-partite delocalization in some $n$-body system, one must generate $(k-1)$-partite entangled states and assess the closest states to the $k$-partite entangled region, such that
\begin{equation}
\label{Ek}
E_k\left(\rho\right)=\tau_k\left(\rho\right)-\tau_k\left(\sigma\right),
 \end{equation}
 In general  \citep{Scholak2011a} $\tau_k$ can be calculated from
\begin{widetext}
\begin{equation}{\tau_{k}}=\sum^{N-k+1}_{i_0=1}{\rho_{i_0 i_0}}\sum^{N-k+2}_{i_1=i_0+1}{\rho_{i_1 i_1}}\ldots\sum^{N}_{i_{k-1}=i_{k-2}+1}{\rho_{i_{k-1}i_{k-1}}}.
\label{T5}
\end{equation}
\end{widetext}
 \noindent Mixed states detected by these measures can be considered to have genuine $k$-partite entanglement as they are not producible by states with $k-1$-partite entanglement \citep{Guhne2005}.
\\

\noindent The biggest challenge in deriving these measures is finding the correct reference states for a given system and measure. However, now that we have proven that we cannot have any overlapping matrices, this gives the added advantage that the purity of $\sigma$ can also be written in terms of the probabilities of the constituent matrices: $p^2_1+p^2_2+\ldots+ p^2_m=\Tr\left(\sigma^2\right)=\Tr\left(\rho^2\right)$, where each constituent matrix is fully delocalized according to its size constraints. As a result of this, these reference states appear to fall in to three main categories, depending on the size of the system and the level of delocalization being measured.

\noindent For example, for reference states in a system of size $n$, where $k-1\geq n/2$, only two states are needed; one with $k-1$-partite entanglement and one with $(n-k+1)$-partite entanglement. When $n=k$ this second matrix will be a pure, separable state with a single mode occupied. The probabilities for such a system will be the solutions to the quadratic equation $p^2_1+(1-p_1)^2=\Tr\left(\rho^2\right)$, such that $p_1=\frac{1}{2}\left(1+\sqrt{2\left({\Tr\left({\rho }^2\right)}\right)-1}\right)$, just like in our previous example with equation \ref{prob2ev}.

\noindent The next category of reference states is where $k-1$ divides $q$ times into system size $n$. Given that each constituent state has the same delocalization ($k-1$) as $\sigma_1$, we will introduce $\sigma_2$ to $\sigma_q$ with equal probability. Therefore the probabilities will again come from a solution to a quadratic equation:
 \begin{equation}
p^2_1+(q-1)\left(\frac{1-p_1}{q-1}\right)^2=\Tr\left(\rho^2\right)
\end{equation}
where we find $p_1=1/q\left(\sqrt{(q^2-q)\Tr\left(\rho^2\right)-q+1}+1\right)$ and $p_2=p_3\ldots=p_q=1/q(1-p_1)$. \\

\noindent An example of these reference states can be seen in equation \ref{sig3}.

\subsubsection{Tripartite entanglement in a 5-body state}
\noindent Now lets look at the final category of reference states: when $k-1$ divides $q$ times with some remainder $r$. This requires $q$ states with $k-1$ mode delocalization as well as one state with $r$ mode delocalization. This time the probabilities are the solution to the quadratic equation:
 \begin{equation}
p^2_1+(q-1)\left(\frac{1-p_1-p_r}{q-1}\right)^2+p_r^2=\Tr\left(\rho^2\right)
\end{equation}
 The solutions being \\
 \begin{widetext}
 \begin{eqnarray}
  p_1&=&\frac{1-p_r+\sqrt{(1-q)(1+p_r(q p_r +p_r-2)-q\Tr\left(\rho^2\right))}}{q} \\
p_2&=&\ldots=p_q=(1-p_1-p_r)/(1-q)
 \end{eqnarray}
\end{widetext}

Solving for $p_r$ is significantly harder. Here we detail one approach, with $n=5$ and $k=3$. Firstly, recall that we are trying to detect tripartite entanglement, so we need to maximize $\tau_3(\sigma)$ for a given value of $p_r$.
 \begin{equation}
\frac{d\tau_3(\sigma)}{d{p_r}}=0
\label{deriv}
\end{equation}
Rewriting $\tau_3(\sigma)$ as a function of $p_r$ with $q=2$ we get:
 \begin{equation}
\tau_3(\sigma)=1-\frac{1}{4} \left(1+3 \Tr\left(\rho^2\right)(1+p_r)-3 p_r(4 p_r^2-4 p_r+1)\right)
\end{equation}
Then we take the derivative as in equation \ref{deriv} and find the roots, selecting the root that is zero when $\rho$ is pure.
 \begin{equation}
p_r=\frac{1}{6} \left(2-\sqrt{1+3  \Tr\left(\rho^2\right)}\right)
\end{equation}

However it turns out the solution to $p_r$ is a piecewise function as when $\Tr\left(\rho^2\right)=\frac{3}{7}$ the solutions to $p_1$ and $p_2$ become complex. Therefore at that point $p_r$ switches to its lowest possible value, $p_r=\frac{1}{3}\left(1-\sqrt{6\left(\Tr\left({\rho}^2\right)\right)-2}\right)$.\\

We are left with the solutions
\begin{eqnarray}
p_1&=&\frac{1}{2} \left(1-p_r+\sqrt{+2 \Tr\left(\rho^2\right)-1 +2p_r-3 p_r^2}\right)\nonumber\\
p_2&=&1-p_1-p_r\nonumber
\end{eqnarray}
 \begin{displaymath}
   p_r = \left\{
     \begin{array}{lr}
       \frac{1}{6} \left(2-\sqrt{1+3  \Tr\left(\rho^2\right)}\right) & : 3/7 \leq \Tr\left(\rho^2\right) \leq 1 \\
       \frac{1}{3}\left(1-\sqrt{6\left(\Tr\left({\rho }^2\right)\right)-2}\right)  & : 9/25 \leq \Tr\left(\rho^2\right) < 3/7
     \end{array}
   \right.
\end{displaymath}

with
$\sigma=p_1{\sigma_1}+p_2{\sigma_2}+p_r{{\sigma_r}}$,
where $\sigma_1$ is a state with modes 1 and 2 fully entangled, $\sigma_2$ is a state with modes 3 and 4 fully entangled, and $\sigma_r$ is a separable state with mode 5 fully populated.
%$\sigma=p_1{\left|W_{12}\right>\left<W_{12}\right|}+p_2{\left|W_{34}\right>\left<W_{34}\right|}+p_r{\left|W_{5}\right>\left<W_{5}\right|}$.
Finally our measure of tripartite delocalization is
\begin{equation}
E_3(\rho)=\tau_3(\rho)-\tau_3(\sigma).
\end{equation}
 \\

In the event of being unable to determine $p_r$, a sufficiently large number of randomly generated reference states will create a distinct region like in figure \ref{E3plot}, along the border of which a curve can be fitted and used in lieu of the exact reference states.\\

%\subsection{General Method}
%
%
%%In order to maximize their $k-1$ entanglement, the number of constituent states in the final mixed state must be minimized.
%
%\noindent If $k-1\geq n/2$ only two states are needed; one with $k-1$-partite entanglement and one with $(n-k+1)$-partite entanglement. Their probabilities are determined just like in our previous example with equation \ref{prob2ev}. The second is when $k-1 < n/2$ and $k-1$ divides $q$ times into $n$. Then $q$ constituent states are required, each with $k-1$-partite entanglement and probabilities
%\begin{equation}\label{p1c2}
%p_1=1/q\left(\sqrt{(q^2-q)\Tr\left(\rho^2\right)-q+1}+1\right)
%\end{equation}
% and
% \begin{equation}\label{p2c2}
% p_2=p_3\ldots=p_q=(1-p_1)/(q-1).
% \end{equation}
%\\
%
%\noindent The final category is when $k-1$ divides $q$ times with some remainder $r$. These probabilities are determined in a similar fashion: as a multivariable quadratic equation $p^2_{1}+p^2_{2}+\ldots+p^2_{r}=\Tr\left({\rho }^2\right)$ with
%
%\begin{widetext}
%\begin{eqnarray}
%% \nonumber to remove numbering (before each equation)
%  p_1&=&1/q(\sqrt{(1-q)(1+p_r(q p_r +p_r-2)-q\Tr\left(\rho^2\right))}+1-p_r)\\
%p_2&=&\ldots=p_q=(1-p_1-p_r)/(1-q).
%\end{eqnarray}
% \end{widetext}
%
%Calculating $p_r$ can prove difficult, so we include an example with $n=5$ and $k=3$ in the supplementary information.
%%$p^2_{1} \geq p^2_{2} \ldots \geq p^2_{q+1}$
%\\
\section{Multipartite Evolution}
\noindent As a demonstration of the effectiveness of these measures, we plot in figure \ref{MultiPlot} the evolution of multi-partite delocalization within a coupled $6$-body $2$-level system, undergoing a dephasing evolution. The Hamiltonian and master equation employed describe an example system of a ring of $6$ chromophores in a photosynthetic light harvesting complex coupled to a bath. The initial state chosen is a pure W-state, i.e maximally entangled. The reference states for each measure were fully determined.
\subsection{Hamiltonian and master equation }
\noindent The Hamiltonian employed in our simulation describes a ring of 6 sites with nearest-neighbor coupling. The energy units are $\textrm{cm}^{-1}$.

\begin{equation}H=\left( \begin{array}{cccccc}
12500 & 300 & 0 & 0 & 0 & 300\\
300 & 12000 & 300 & 0 & 0 & 0\\
0 & 300 & 12500 & 300 & 0 & 0\\
0 & 0 & 300 & 12000 & 300 & 0\\
0 & 0 & 0 & 300 & 12500 & 300\\
300 & 0 & 0 & 0 & 300 & 12000\\
\end{array}
\right)
\end{equation}\\

\noindent we use the Redfield equation within the secular approximation \citep{Fassioli2010a}. The density matrix of the system obeys the following master equation:
\begin{equation}\frac{\partial\rho \left(t\right)}{\partial t}=-\rm i\left[H,\rho \left(t\right)\right]+D\left(\rho \left(t\right)\right).\end{equation}

\noindent The first term on the right hand side describes purely coherent evolution and the second induces dephasing and relaxation between excitonic states of the system through the dissipator operator $D\left(\rho \left(t\right)\right)$. The dissipator reads
\begin{widetext}
\begin{equation}
D\left(\rho \left(t\right)\right)=\sum_{\omega }{\sum_{m,n}{\gamma \left(\omega \right)\left[A_n\left(\omega \right)\rho \left(t\right)A^\dag_m\left(\omega \right)-\frac{1}{2}\{A^\dag_m\left(\omega \right)A_n\left(\omega \right),\rho \left(t\right)\}\right]}}
,\end{equation}

\end{widetext}
\noindent where $A_n\left(\omega \right)=\sum_{\epsilon_{k'}-\epsilon_k=\omega }{a^*_n{\left(\phi_k\right)a}_n}\left(\phi_k'\right)\left.|\phi_k\right\rangle \left\langle \phi_k'|\right.$ are the Lindblad operators, with $a_n$ the site coefficients of exciton $\left.|\psi \right\rangle$ such that $\left.|\psi \right\rangle =\sum^N_n{a_n\left.|n\right\rangle }$. We assume that site fluctuations are independent. The rates $\gamma(\omega)$ are given by $\gamma_{mn}(\omega) \equiv \gamma(\omega)=2\pi J\left(\left|\omega \right|\right)\left|N\left(-\omega \right)\right|$.
$J(\omega)$ is the spectral density characterizing the system-phonon coupling, which we assume to be ohmic with Drude cutoff, i.e. $J\left(\omega \right)=2{E_r}{\omega_c}{\omega}/\pi\left({\omega_c}^2+\omega^2 \right)$
, where $E_r$ is the reorganization energy, $\omega_c$ is the cutoff frequency and  $N\left(\omega \right)$ is the thermal occupation number. In this simulation we chose a value of $300\ \textrm{cm}^{-1}$ for $E_r$ equal to the level of coupling within the system. The temperature chosen was $77\ \textrm{K}$.

\subsection{Reference states and analysis}

\begin{figure}[htbp]
  % <PostScript inclusion special>
\begin{center}
\includegraphics[scale=0.45]{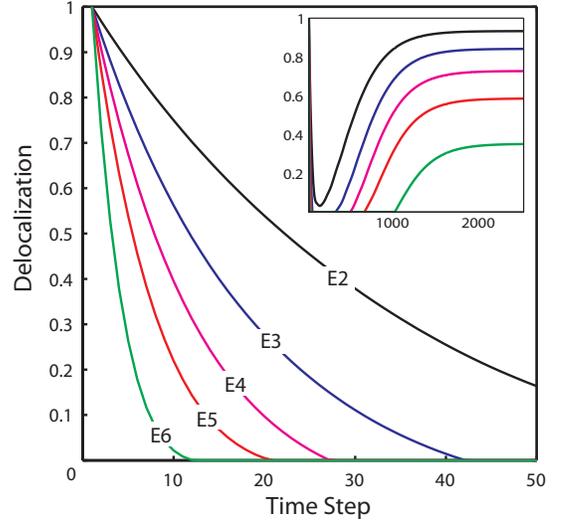}
\end{center}
\caption{
Early dephased evolution of multipartite delocalization in a 6-body system over the first 50 time steps. The inset shows the long term evolution of multipartite delocalization. All measures have been normalized.}
\label{MultiPlot}
\end{figure}

For $E_3$ we derived the reference state $\sigma_3=p_1{\left|W_{12}\right>\left<W_{12}\right|}+p_2{\left|W_{34}\right>\left<W_{34}\right|}+p_3{\left|W_{56}\right>\left<W_{56}\right|}$, where\begin{eqnarray}\label{sig3}
\left|W_{12}\right>&=&\frac{1}{\sqrt{2}}\left({\left|100000\right>}+{\left|010000\right>}\right) \equiv \frac{1}{\sqrt{2}}\left({\left|1\right>}+{\left|2\right>}\right) \nonumber\\ \left|W_{34}\right>&=&\frac{1}{\sqrt{2}}\left(\left|3\right>+{\left|4\right>}\right)  \nonumber\\  \left|W_{56}\right>&=&\frac{1}{\sqrt{2}}\left(\left|5\right>+{\left|6\right>}\right).  \nonumber\\
 \end{eqnarray}
 The probabilities for $\sigma_3$ were calculated as
 \begin{equation}
p_1=\frac{1}{3}\left(\sqrt{6\left(\Tr\left({\rho }^2\right)\right)-2}+1\right)  \nonumber\\ \textrm{and}
 \ p_2=p_3=\frac{1}{2}\left(1-p_1\right)\nonumber\\
   \end{equation}\\
\noindent For $E_4\equiv \tau_{4}\left(\rho\right)-\tau_{4}\left(\sigma_4\right)$, our reference state is $\sigma_4=p_1{\left|W_{123}\right>\left<W_{123}\right|}+p_2{\left|W_{456}\right>\left<W_{456}\right|}$, and
 \begin{eqnarray}
\left|W_{123}\right>&=&\frac{1}{\sqrt{3}}\left(\left|1\right>+\left|2\right>+\left|3\right>\right)\nonumber\\
\left|W_{456}\right>&=&\frac{1}{\sqrt{3}}\left(\left|4\right>+\left|5\right>+\left|6\right>\right).\nonumber\\
 \end{eqnarray}\\
\noindent For $E_5\equiv \tau_{5}\left(\rho\right)-\tau_{5}\left(\sigma_5\right)$, our reference state is
$\sigma_5=p_1{\left|W_{1234}\right>\left<W_{1234}\right|}+p_2{\left|W_{56}\right>\left<W_{56}\right|}$, and
 \begin{eqnarray}
\left|W_{1234}\right>&=&\frac{1}{\sqrt{4}}\left(\left|1\right>+\left|2\right>+\left|3\right>+\left|4\right>\right)\nonumber\\
\left|W_{56}\right>&=&\frac{1}{\sqrt{2}}\left(\left|5\right>+\left|6\right>\right).\nonumber\\
 \end{eqnarray}\\
\noindent Finally for $E_6\equiv\tau_{6}\left(\rho\right)-\tau_{6}\left(\sigma_6\right)$, our reference state is
$\sigma_6=p_1{\left|W_{12345}\right>\left<W_{12345}\right|}+p_2{\left|W_{6}\right>\left<W_6\right|}$, and

 \begin{eqnarray}
\left|W_{12345}\right>&=&\frac{1}{\sqrt{5}}\left(\left|1\right>+\left|2\right>+\left|3\right>+\left|4\right>+\left|5\right>\right)\nonumber\\
\left|W_6\right>&=&\left|6\right>.\nonumber\\
 \end{eqnarray}

\noindent As there are only two constituent states in the reference states of measures $E4$ to $E6$, the probabilities are the same as equation \ref{prob2ev}. In general the procedure is to maximize the delocalization in each constituent state, and weight those states accordingly with ranked probabilities, such that $p_1 \geq p_2 \geq\ldots \geq\ p_{m}$.\\

\noindent The initial state chosen for figure \ref{MultiPlot} is a non-stationary state of the system Hamiltonian. Thus, under short term evolution we observe the delocalization decay smoothly from a fully entangled state down to near-zero, with higher orders of delocalization disappearing in order. The long term evolution of the state shows an increase in multipartite delocalization as the system enters a steady state and the majority of the excitation lies in the lowest energy eigenstate of the system. This increase in multipartite delocalization shouldn't be surprising, as the lowest energy eigenstate is highly delocalized in the basis we have chosen.\\
\section{Conclusions}
\noindent In this paper we have demonstrated, by using only the purity and statistical moments, that one can analytically distinguish mixed states with different, quantifiable levels of entanglement in the single excitation subspace. We have taken advantage of the concept of tiered separability in deriving analytical measures of multipartite delocalization. By construction, these measures decrease under loss of information, meaning their convexity need not be proven. The key idea of our approach is to calculate the distance from a state to the next closest $(k-1)$-body entangled state with the same level of purity. Rather than minimizing the distance over a set of randomly generated reference states, the reference states are carefully selected. This allows for instant detection of delocalization, for any level of separability, at any level of decoherence.\\

 \begin{acknowledgments}
This work was supported by the Natural Sciences and Engineering Research Council of Canada, DARPA (QuBE)
and the United States Air Force Office of Scientific Research(FA9550-13-1-0005) to G.D.S. C.S thanks Nicolas Quesada and Aurelia Chenu for fruitful discussions. We also thank Florian Mintert for his vital input on proving the reference states.
   \end{acknowledgments}

\bibliography{MixedMultipartitePaper}
\end{document}